# The predictability of the Hirsch index evolution

Michael Schreiber

*schreiber@physik.tu-chemnitz.de*
Institute of Physics, Chemnitz University of Technology, 09107 Chemnitz (Germany)

**Abstract**
The h-index can be used as a predictor of itself. However, the evolution of the h-index with time is shown in the present investigation to be dominated for several years by citations to previous publications rather than by new scientific achievements. This inert behaviour of the h-index raises questions, whether the h-index can be used profitably in academic appointment processes or for the allocation of research resources.

**Introduction**
Due to its simplicity the so-called Hirsch index or h-index has become attractive as a frequently used metric for describing the scientific achievements of a researcher. It was introduced by Hirsch (2005) as the largest number *h* of publications of a scientist which have received at least *h* citations each. This means $h = \max\{r \mid c(r) \geq r\}$. Here $c(r)$ denotes the number of citations to the paper at rank *r*, after the papers have been sorted according to decreasing $c(r)$.

In spite of several weaknesses and in spite of doubts how representative this measure is, the h-index has become popular and is often used in academic appointment processes and evaluation procedures of research projects. Hirsch (2007) determined a high correlation comparing the h-index values of researchers after 12 years and after 24 years of their careers. He concluded that "the h-index has the highest abibility to predict future scientific achievement".

Utilizing a complicated fit with 18 parameters, Acuna, Allesina, and Kording (2012) were able to predict the future h-index rather accurately for several years. It is controversial (Rousseau and Hu, 2012) whether such an approach is meaningful. In a validation study Garcia-Perez (2013) showed that the predicted h-index trajectories overestimate the future h indices more and more for later and later target years. I have recently demonstrated (Schreiber, 2013) that the increase of the h-index with time after a given point of time (e.g., the appointment year or the evaluation year) is for several years not related very much to the scientific achievements after that date. Rather the growth of the h-index is nearly the same for several years, irrespective of whether further work had been published or not after that date.

In my previous publication I have presented examples for the rather smooth increase of the h-index with time thus confirming that the h-index is a good predictor of itself. However, I have also presented evidence in 4 examples that the growth of the h-index does not depend very much on the factual performance for several years in the future but rather results mostly from previous, often rather old publications. As evidence I had selected the most impressive example years so that the deviations were small for a particularly long time interval and not representative. It is the purpose of the present investigation to analyze quantitatively the duration for which the h-index remains unchanged or only slightly changed.

Figure 1: The Hirsch index $h = h_y(y)$ for the publication record of the present author (top line). The dependence of $h_s(y)$ is shown for $s$ = 1976, 1977, ... (from bottom to top). For the years $s$ = 1980, 1990, and 2000 thick broken lines are utilized. $y$ is the year of evaluation, $s$ is the last year from which publications are taken into consideration.

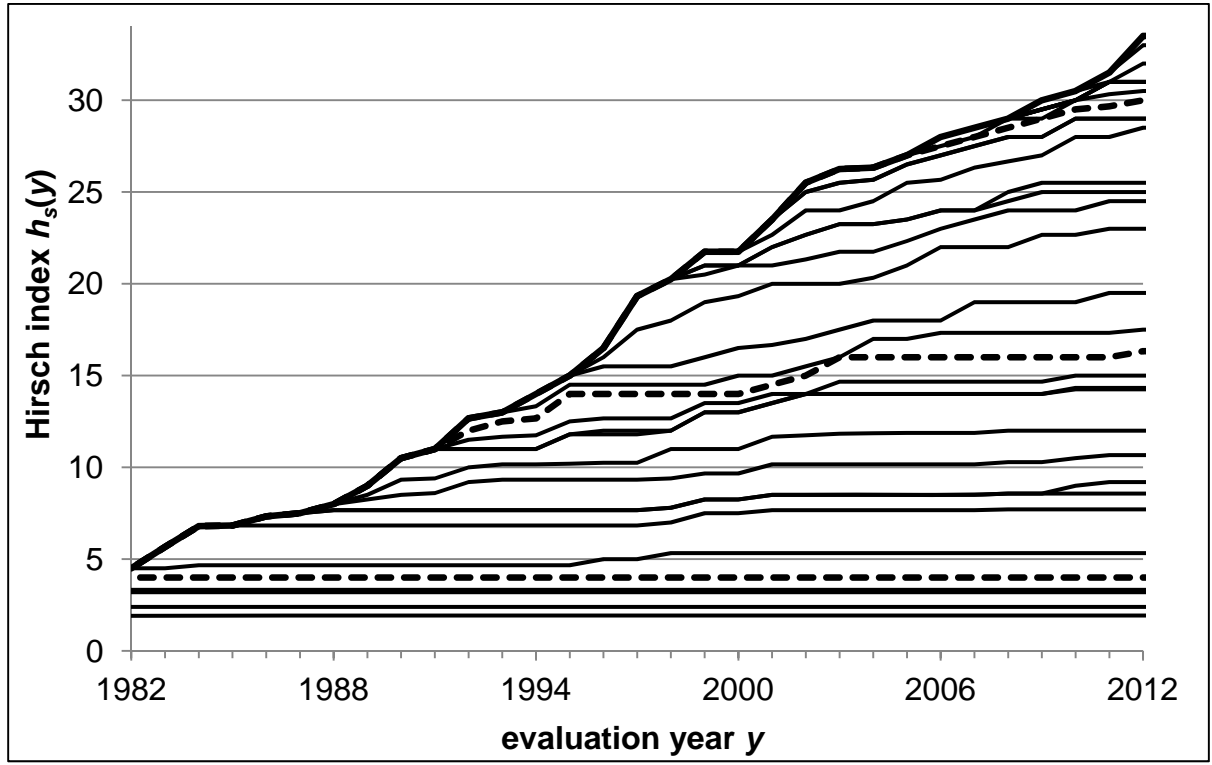

**The citation data and the calculation of the h-index**

I harvested my own citation record from the ISI Web of Science database in March 2013. I determined the citations up to a given year $y$ and counted the publications with high citation frequencies selectively, namely considering only publications up to a certain year $s \leq y$. This yields the selective h-index $h_s(y)$ for the year $y$, under the assumption that I had stopped publishing in year $s$. Of course, if $s = y$ then the usual h-index $h = h_y(y)$ is obtained. As $h$ and likewise $h_s(y)$ are restricted to integer values, a graphical representation of $h_s(y)$ curves is very difficult to survey, because many values coincide. Therefore I had restricted by previous investigation (Schreiber, 2013) to selective values of $s$ and discussed the resulting curves qualitatively.

I have repeated the analysis now with the updated data for the interpolated version of the h-index, which is obtained after a piecewise linear interpolation of the citation distribution $c(r)$ between $r$ and $r + 1$, as suggested and utilized previously by Rousseau (2006), van Eck and Waltman (2008) and myself (Schreiber, 2008, 2009). The interpolated h-index then results from the solution of $c(h) = h$. By truncating the interpolated index values one obtains the usual integer results.

In Fig. 1 the determined $h_s(y)$ curves are presented, showing the expected rather smooth behaviour. The original h-index $h = h_y(y)$ shows a steady increase with a slope of approximately one index point per year. Selecting a particular year $s$, one can see that for a duration of several years $t = y - s$ the deviation $d = h_y(y) - h_s(y)$ remains rather small, indicating that the growth of the h-index after the year $s$ is mostly due to the publications until the year $s$. In fact, for all reasonable values of $s$ (values $s > y$ are not meaningful) the initial part of the curves $h_s(y)$ in Fig. 1 cannot be observed, because the values coincide with $h_y(y)$. Of course, the $h_s(y)$ curves level off for recent years, because the very old publications are either highly cited and belong already to the h-defining set of papers for several years, or they are so lowly cited that they have small chance to become relevant for the h-index. Nevertheless, there are some exceptions from this argumentation which lead to the prominent increases that some of the curves feature in Fig. 1. Furthermore, there is always the possibility of so-called sleeping beauties which have not received a significant number of citations for a long time, but then suddenly are cited frequently. This has happened to some of my papers about quasicrystals, because the subject which was a hot topic in the nineties has become topical again after the Nobel prize in chemistry was awarded to D. Shechtman in 2011.

**The inert behaviour of the h-index**
In order to obtain a more quantitative description of the observed behaviour, I have determined the time span $t_0$ for which the h-index does not differ, irrespective of whether I had performed as I did or whether I had not published any further work after the selected year $s$. The results are shown in Fig. 2. The bottom line indicates that for most values of $s$ the h-index would not have changed for the next 3 years. Between 1982 and 2008 this inert behaviour is observed on average for 3.3 years, i.e. $\bar{t}_0 = 3.3$, where the overbar denotes the average. I have not included more recent years, because for $s = 2008$ I get $t_0 = 4$, so that $y = s + t_0 = 2012$ equals already the last year covered by the dataset. Therefore for more recent years $s$ the restriction $t_0 < 2012 - s$ effectively limits $t_0$ and it is quite likely that $t_0$ will grow further in the future, i.e. for larger values of $y$. In principle this could even happen already for $s = 2008$. I have also excluded the first six years from the average in order to avoid possible problems with a transient behaviour of the index at the beginning of my career. However, including this initial period, the average duration of coinciding $h_y(y)$ and $h_s(y)$ values would change only slightly to $\bar{t}_0 = 2.9$.

Figure 2: Time span $t_d = y - s$ for which the h-index $h_s(y)$ would have remained the same ($d = 0$) or deviated at most by $d = 1, 2,$ or $4$ (from bottom to top) index points from the factual index values $h_y(y)$, if I had stopped publishing in the selected year $s$. The thin broken line indicates the border $t_d = 2012 - s$, which limits the curves for $y = 2012$.

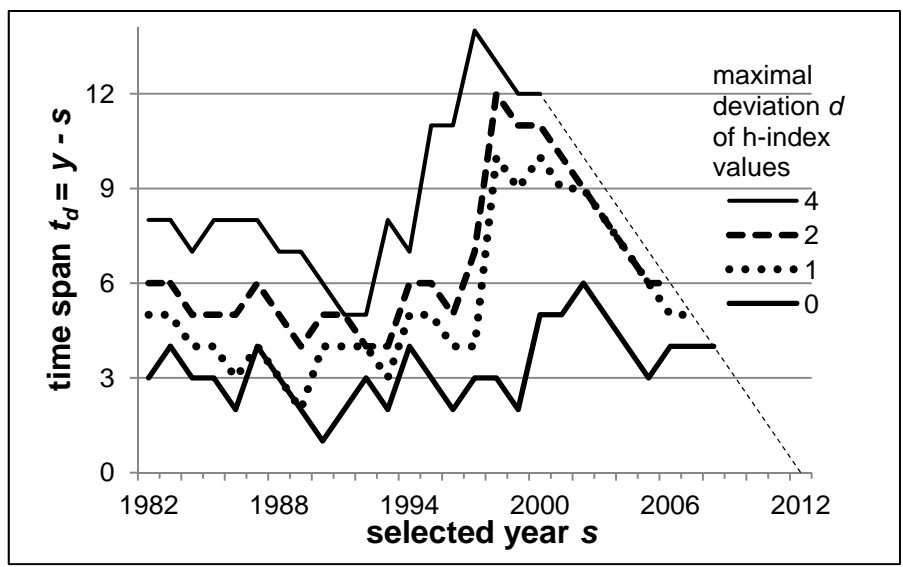

It is well known, that the accuracy of the h-index does not allow a meaningful distinction between researchers with nearly the same index values. In other words, small differences are not meaningful. Therefore I have also calculated and included in Fig. 2 the duration times $t_d$ for which my values of $h_y(y)$ and $h_s(y)$ differ by not more than $d = 1, 2,$ or $4$ index points. It is not surprising that the obtained values of $t_d(s)$ are considerably larger than $t_0(s)$. On average I obtained $\overline{t_1} = 5.4$, $\overline{t_2} = 6.6$, and $\overline{t_4} = 8.7$, excluding again values for $s \leq 1981$ and for $s \geq 2008$, 2007, 2001, respectively, because the border line $y = s + t_d = 2012$ was already reached. Again, including the early years does not change the averages much, in this case I determined $\overline{t_1} = 4.8$, $\overline{t_2} = 5.9$, $\overline{t_4} = 8.3$.

The decrease of the curves in Fig. 2 around 1990 indicates that rather recent publications influence the h-index evolution. The prominent increases after 1994 show that now the h-index is dominated by past scientific achievements. Of course with increasing values of $h$ it is more and more difficult for new publications to obtain the number of citations necessary for becoming relevant for the h-index. The distinct drop of the curves on the right-hand side of Fig. 2 shows that now rather recent publications have had an effect. In fact, a closer investigation of my citation record revealed that this is due to my new hobbyhorse: My investigations on the h-index and other topics in Scientometrics have quickly made it into the set of $h$-defining publications.

**The predictability of the h-index**
Another possibility to describe the inert behaviour or the predictability of the h-index in a quantitative way is to consider a certain year $y$ and to determine for different time spans $t = y - s$ the deviation $d_t = h_y(y) - h_s(y)$, which is caused by excluding all papers after the selected year $s$ from the evaluation of the h-index. Respective results are presented in Fig. 3. It can be seen that for a time span of $t = 3$ years the deviation is small, often there is no difference at all as already observed in Fig. 2. The average deviation since $y = 1982$ is $\overline{d_3} = 0.31$. Excluding the publications for the last $t = 4$ years, somewhat larger deviations can be found but they are still small, on average $\overline{d_4} = 0.75$. Even for $t = 6$ years, the average difference is only $\overline{d_6} = 2.1$, which most people would probably consider as not being relevant. For $t = 8$ years of unproductivity, the average influence on the h-index is $\overline{d_8} = 3.5$.

Figure 3: Difference $d_t = h_y(y) - h_s(y)$ for $t = 3, 4, 6, 8$ (from bottom to top),
if the last $t = y - s$ years are excluded from the determination of $h$ in the year $y$,
i.e., if I had stopped publishing in the selected year $s = y - t$.
For $t = 8$ the first two data points are missing, because my publication record starts
in the year 1976 and these missing data points correspond to $s = 1974$ and $1975$.

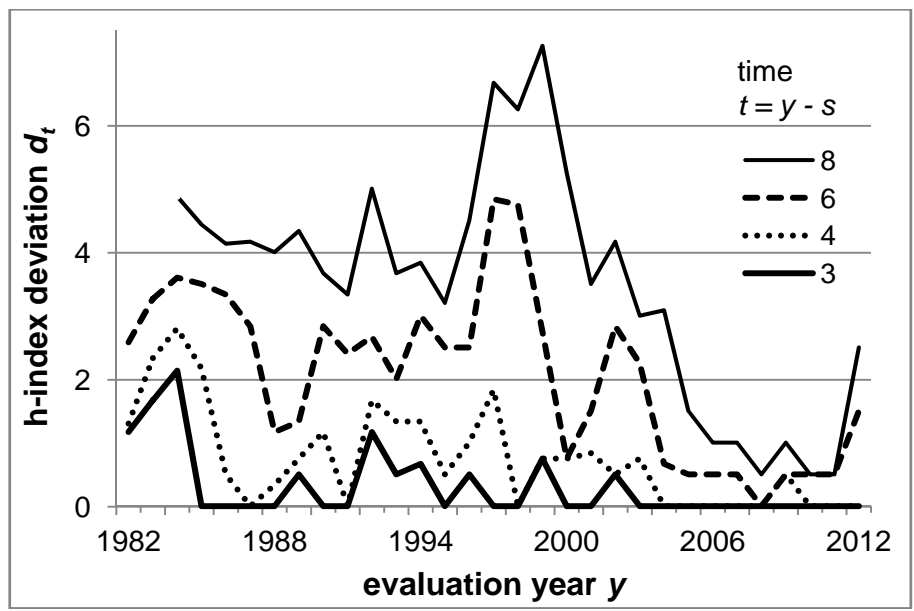

For individual years there are some outliers in the curves presented in Fig. 3. The rather large deviations in the late nineties correspond to the low values of $t_d$ in Fig. 2 around 1990 and reflect the high impact of several publications around 1990 which have quickly made an effect on my h-index and then dominated its evolution in the late nineties. Likewise the small values of $d_t$ before and around 2010 correspond to the high values of $t_d$ in the late nineties in Fig. 2, and can be likewise explained by rather old publications dominating the h-index evolution.

For a comparison between Figs. 2 and 3 it should be noted that the curves in Fig. 2 are plotted in dependence on $s$, while the curves in Fig. 3 are presented as a function of $y$. Thus there is a tilt of $t = y - s$ years between corresponding features in these two figures to take into consideration.

**Discussion**
It will be interesting to see whether other citations records show a similar behaviour, not only the overall inert behaviour, but also whether the predictability can be likewise attributed predominantly to relatively old publications. Therefore further studies of the persistence of the h-index values in terms of the duration $t$ should be performed. It would also be interesting to see whether such features as discussed above, namely distinct maxima or minima and prominent increases or decreases of the curves as in Figs. 2 and 3 can be found and attributed to specific details of other citation records.

In spite of the discussed inert behaviour the obtained results corroborate the predictive power of the h-index (Hirsch, 2007). It is tempting to assume that a predicted growth of the h-index can be correlated with the future performance of a candidate. This would make the h-index a possibly useful measure in academic appointment processes and for the allocation of research resources. However, as the present investigation has shown, the future development of the h-index is dominated for several years by citations to previous publications. This means that a high h-index value of a candidate can be expected to increase after this person is hired, even if he or she goes to sleep after the appointment and does not publish any further work. On the other hand, the past evolution of the h-index does not automatically mean that the candidate has performed good work in recent years. The h-index would most likely have grown more or less as it did, even if the candidate had gone to sleep several years ago. In conclusion, the present investigation raises doubts about the usefulness of the h-index for predicting future scientific achievements.

On the other hand, the observed inert behaviour of the h-index bears testimony of the significance of a researcher's past achievements. This is certainly an aspect which should be taken into account also in appointment processes or for the purpose of allocating resources.